# Ionized particle transport in reactive HiPIMS discharge: correlation between the energy distribution functions of neutral and ionized atoms

A el Farsy[1,2,*], D Boivin[1], C Noel[1], R Hugon[1], S Cuynet[1], J Bougdira[1], L de Poucques[1]
[1] Université de Lorraine, CNRS, IJL, F-54000 Nancy, France.
[2] Université Paris-Saclay, CNRS, Laboratoire de Physique des Gaz et des Plasmas, 91405 Orsay, France.

*Corresponding author: *abderzak.el-farsy@universite-paris-saclay.com*

**Abstract:**

We investigated the transport titanium ions produced in a reactive high-power impulse magnetron sputtering (HiPIMS) device used for TiN coating deposition. Time-resolved mass spectrometry measurements of ionized sputtered atoms correlated to time-resolved tuneable diode-laser induced fluorescence (TR-TDLIF) measurements of neutral sputtered atoms were used to understand transport features. Based on ion energy distributions of $Ti^+$, we identified four populations of ions and explore their physical origins. The signals of all ion populations decrease strongly when only 1% $N_2$ is added to the $Ar/N_2$ gas mixture. Time resolved mass spectrometry confirms the result reported in previous work: the fast target poisoning when nitrogen is added in HiPIMS discharges. Based on the measured energy distribution functions of $Ti^{++}$, $N^+$, $N_2^+$, and $Ar^+$, we discuss the production of these ions in HiPIMS discharges. The temperature of thermalized sputtered neutral atoms determined by previous TR-TDLIF measurements evidences the physical origin of an ion population with energies lower than 4 eV. According to discharge pressure, cathode voltage, and ion type, we also discuss the physical origins of high-energy ions (>4 eV).

# I. Introduction

Reactive high-power impulse magnetron sputtering (R-HiPIMS) is an ionized physical vapor deposition method [1,2] in which high power is applied during short pulses to achieve greater than 50% ionization of the metallic sputtered vapor. The power is turned on during short pulses and turned off for longer periods post-discharge to avoid overheating the magnetron cathode and arc formation. Sputtered atoms are ionized to control their energy on the substrate and therefore to monitor the properties of the thin film. Indeed, ion energy is influenced by the electric field produced when a negative bias is applied to the substrate. Ion-assisted film growth meets the requirements of applications for which it is necessary to improve the quality of the deposited film, for example, in terms of mechanical properties (dense films with reduced porosity, better corrosion resistance) and in the case of substrates with complex 3D geometries. Therefore, with high spatiotemporal knowledge of the properties of the incoming film-forming species (neutral atoms and ions) in R-HiPIMS, the physical properties of the film can be adjusted to target applications [3–5]. In other words, a clear understanding of the basic processes governing the transport of neutral atoms and ions is crucial for optimizing material deposition processes for new applications.

In the sputtering process, the transport of neutral atoms of metallic vapor has been thoroughly investigated for different target materials using optical diagnostics, mostly tunable diode-laser induced fluorescence (TDLIF) or laser absorption spectroscopy [6–13]. Because R-HiPIMS is performed at low pressure ($p \approx 1$ Pa), an intermediate transport regime, not purely ballistic nor diffusive, has been highlighted in the cases of sputtering titanium and tungsten targets [14,15]. In these two cases, and thanks to specific laser diodes, a diagnostic has been developed to perform time-resolved measurements of neutral atoms. However, no diode laser is available at wavelengths allowing the measurement of ion velocity distribution functions. Therefore, ion energies in HiPIMS processes are commonly characterized by mass spectrometry [16,17].

The study of metallic ion transport in HiPIMS plasmas is much more complicated than the case of the transport of neutral atoms for three reasons. First, ions are produced from neutral atoms sputtered from the target, which have an initial velocity distribution (the Thompson distribution [18,19]) that is modified by collisions with the buffer gas during transport. Second, because ions are strongly driven by the potential/electric fields, any variation of the cathode voltage and/or plasma potential impacts their transport. Third, due to magnetized plasma discharge in an $\vec{E} \times \vec{B}$ configuration, the emergence of structures called 'spokes' can move the charged particles in the azimuthal direction [20–23], which can also importantly affect ion transport in the axial direction. Although numerous studies have measured ionic fluxes by mass spectrometry [17,24–27], in most cases, the long discharge times used ($t_d > 30$ μs) resulted in the simultaneous involvement of sputtering and transport processes. Palmucci et al [17] distinguished only two ion populations (cold and hot ions at 0–5 and 5–30 eV, respectively) despite measurements performed at a temporal resolution of 2 μs and the use of short HiPIMS pulses of 5 μs.

In this work, we study ion transport in a short-pulse R-HiPIMS process used for titanium nitride (TiN) thin-film deposition. TiN film was chosen for its good mechanical and optical properties, excellent chemical resistance, and good biocompatibility, but we note that film properties are not the specific focus of this work. In our previous work on the transport of sputtered neutral titanium atoms in Ar/$N_2$ HiPIMS discharge, we observed that the deposited flux and energy remain constant with increasing $N_2$ content (≥1%) in the gas mixture [14]. This surprising behaviour with increasing fraction of reactive gas raises the need to characterize the ions by time-resolved mass spectrometry. Moreover, the correlation between time-resolved TDLIF (TR-TDLIF) [12] and mass spectrometry measurements can provide a clear understanding of the transport properties of the different ions produced in the HiPIMS process.

In the first part of this work, we characterize the effects of nitrogen on the ionization of sputtered atoms (i.e., the production of $Ti^+$ ions). Our analysis of ion energy distribution functions (IEDFs) by time-resolved mass spectrometry revealed four populations of $Ti^+$ ions. In the second part, we characterize the IEDFs of different types of ions ($Ti^{++}$, $N^+$, $N_2^+$, and $Ar^+$). In the last part, we report experiments performed over a wide range of discharge parameters (pressure, distance between the target and the orifice of the mass spectrometer, gas mixture composition) to investigate the physical origins of the different ion populations. This investigation relies on the correlation between the IEDFs to the previous measurements of the flux of neutral titanium atoms performed by TR-TDLIF and the measurements of cathode voltage in which was performed using a voltage probe.

## II. Experimental setup

The experiments were conducted in the reactor used in our previous work [6], comprising a cylindrical stainless-steel chamber 35 cm in height and 30 cm in diameter and equipped with a turbomolecular pump to achieve a vacuum pressure of $2 \times 10^{-4}$ Pa before injecting gas into the chamber. A water-cooled 2-inch balanced magnetron cathode with a 99.99% purity titanium target 3 mm thick was used as the sputtering source. At a constant total flow rate of 18 sccm, Ar and $N_2$ gases were added into the chamber at various $N_2$/(Ar+$N_2$) flow-rate ratios (0–10%). The gas arriving in the chamber was inserted farther than 20 cm from the magnetron cathode. The discharge was powered by a MELEC SIPP 2000 HiPIMS generator ($U_{max}$ =1 kV, $I_{max\_peak}$ = 200 A). To correlate mass spectrometry measurements with previous TR-TDLIF results [14], the HiPIMS discharge power was fixed at $P = 350$ W cm$^{-2}$ ($U = 500$ V, $I = 20$ A) during discharge ($t_d = 10$ μs) and the repetition frequency was fixed at 1 kHz (corresponding to a duty cycle of 1%). These parameters were specifically chosen to temporally separate the sputtering phase during discharge from the post-discharge transport phase, and are discussed further in [14]. The HiPIMS voltage was monitored by a P6015A Tektronix probe with a bandwidth of 75 MHz and a rise time of 4 ns. The voltage probe was connected to the magnetron within 30 cm from the cathode.

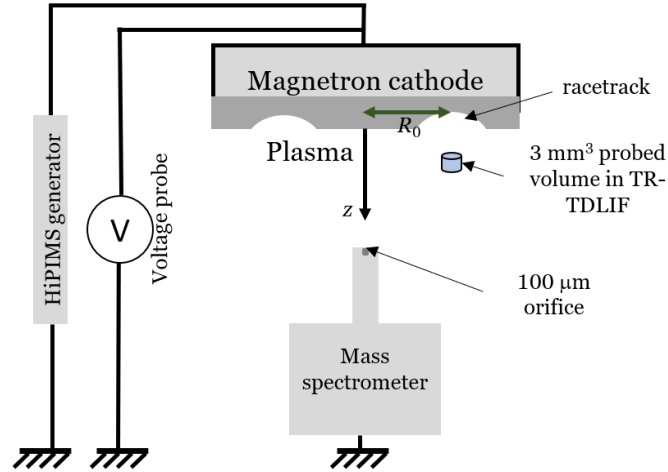

**Figure 1**. Schematic of the experimental setup.

A HIDEN EQP 300 instrument was used to measure the time-resolved IEDFs (TR-IEDFs) of $Ti^+$, $Ar^+$, $N^+$, $Ti^{++}$, and $N^{++}$. The grounded 100-µm-diameter orifice of the mass spectrometer was placed in the virtual position of the substrate. Local TR-TDLIF measurements have been obtained in volumes located above the racetrack ($R_0 = 1.3$ cm) [14]. However, because of the reactor geometry, mass spectrometry measurements were made in front of the centre of the target ($R_0 = 0$ cm, $z = 5$ or 12 cm; see schematic in figure 1). At 0.4 Pa and a high percentage of $N_2$, the plasma is not stable when the head of the mass spectrometer is at $z = 5$ cm. A Tektronix AFG3022C function generator was used to trigger the HiPIMS pulses and synchronize the secondary electron multiplier (SEM) detector of the mass spectrometer with the HiPIMS generator, allowing time-resolved measurements. A 20 µs integration gate was chosen to obtain the best compromise between measured signal intensity (good signal-to-noise ratio over a dwell time of 1 ms) and characterization of the full HiPIMS period (discharge + post-discharge) in a reasonable time. In addition, because the SEM detector is synchronized with the plasma pulses, an ion's detection time at the SEM corresponds to its arrival time at the orifice plus its transit time in the spectrometer (time of flight, TOF). Therefore, measurement corrections account for ion-specific TOFs inside the mass spectrometer, which are determined by the ions' velocities in each part of the spectrometer (e.g., extractor, analysers, detector) and are obtained by different accelerating or decelerating voltages. The corresponding TOF values for the ions of interest in our mass spectrometer were calculated according to the formula provided by the manufacturer (table 1). These transit times strongly depend on the ions' mass/charge ratios, but are not sensitive to their energy at the entrance of the mass spectrometer for energies between 0 and 100 eV [26]. The quantitative calibration of a mass spectrometer is not an easy task because the transmission of ions depends on multiple parameters: orifice, extractor, analysers, detector response, and ionic mass and energy. Here, our mass spectrometer measurements were not calibrated, so we discuss only qualitative signal variations.

**Table 1.** Times of flight (TOF) for different ions in the HIDEN EQP 300 mass spectrometer.

| Ion | Mass/charge | TOF (μs) |
|---|---|---|
| $Ti^+$ | 48 | 83 |
| $Ti^{++}$ | 24 | 58 |
| $Ar^+$ | 40 | 76 |
| $N^+$ | 14 | 44 |
| $N_2^+$ | 28 | 63 |

The entire HiPIMS period was scanned using a variable time step between each measurement. We started with a time step $\delta t = 20$ μs until the newly created ions appear (ions of 20 eV in figure 2). Then, we used shorter step of $\delta t = 5$ μs to better describe the temporal evolution of this new energy distribution until the mass spectrometer signal decreases to about 50% of the maximum observed value. Then, $\delta t$ was gradually increased until the end of the post-discharge period to track later slight temporal variations of the IEDF. Using this specific method, one temporal evolution is characterized by approximately 25 measurements, which takes about 60 min. Figure 2 illustrates TR-IEDFs for $Ti^+$ ions under a typical plasma condition (0.7 Pa, $z = 5$ cm, pure Ar). The ions in this figure can be classified into three groups based on their energies: ions of the first, second, and third groups have energies of 0–15, 15–50, and >50 eV, respectively. Here and in all following figures, the time indicated corresponds to the ions' arrivals at the mass spectrometer orifice, and the origin time ($t = 0$ μs) corresponds to the start of the HiPIMS discharge.

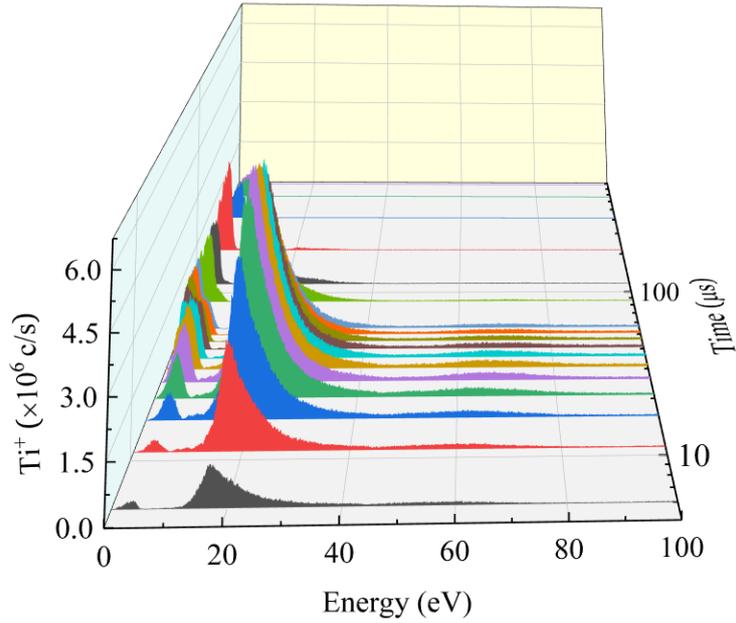

**Figure 2.** Typical measurement of the TR-IEDF of $Ti^+$ in the HiPIMS process ($P_{HiPIMS} = 350$ W cm$^{-2}$, $p = 0.7$ Pa, pure Ar, $z = 5$ cm).

## III. Results and discussion

### III.1. Reactive gas effect on ionization of sputtered atoms

### III.1.1. Identification of four ion populations

At 0.7 Pa, one can observe the existence of three groups of Ti$^+$ ions (figure 2). However, by increasing the working pressure to 2.7 Pa, four ion populations can be identified (figure 3):

- Population 1 (*Pop1*) has energies of about 0–4 eV. This population appears late during the post-discharge period, ~40 µs after the beginning of the discharge. At the studied conditions, this population is smaller than the other populations and is difficult to identify at low pressure (0.7 Pa).

- Population 2 (*Pop2*) has energies of 4–15 eV; it also depends on the working pressure, and is less pronounced at 0.7 Pa.

- Population 3 (*Pop3*) has energies of 15–50 eV with a peak around 20 eV. This population is predominant at low pressure (0.7 Pa) but is still pronounced at higher pressure (2.7 Pa).

- Population 4 (*Pop4*) includes ions with energies >50 eV and behaves similar to *Pop2*, i.e., it becomes more pronounced with increasing pressure.

These populations are defined purely in terms of ion energy and their possible physical origins will be discussed in section III.3.

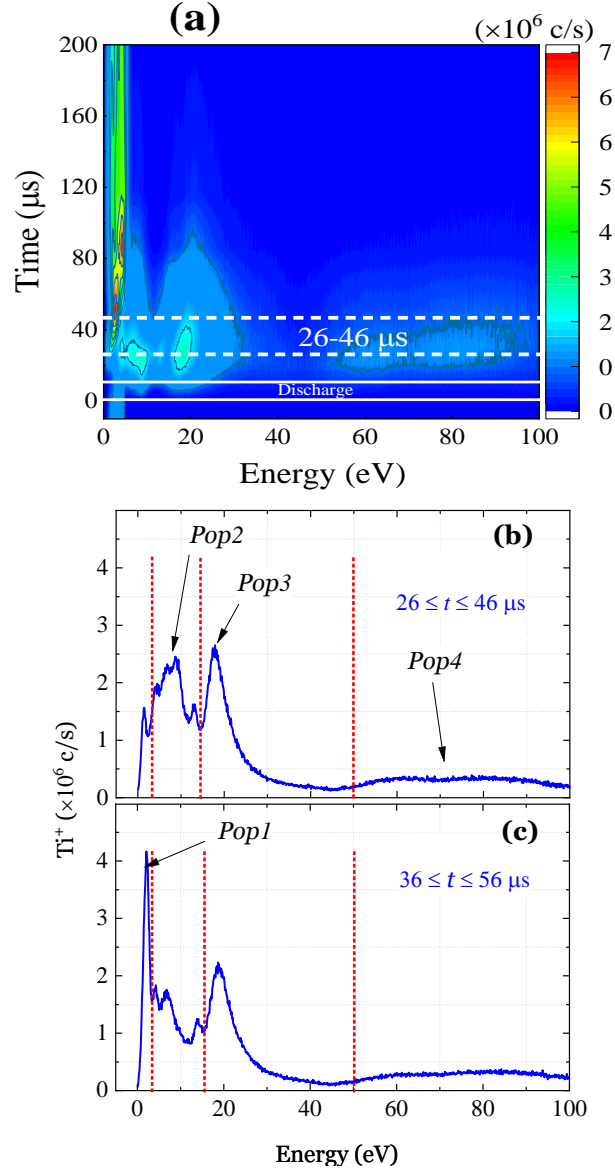

**Figure 3.** (a) Ti$^+$ TR-IEDF measured at $P_{\text{HiPIMS}}$ = 350 W cm$^{-2}$, $p$ = 2.7 Pa, pure Ar, and $z$ = 5 cm. IEDFs obtained at (b) $t$ = 26–46 µs (delimited by the dashed lines in (a)) and (c) $t$ = 36–56 µs after the start of the discharge.

From the measured TR-IEDFs, we calculated the ion flux (the area under the curve) of each population as

$$\text{ion flux}_{\text{popX}}(t) = \int_{E_{\text{l popX}}}^{E_{\text{u popX}}} \text{signal}(t, E) \, dE, \qquad (1)$$

where $E_l$ and $E_u$ are the lower and upper energy limits of the considered population ('popX'), respectively. We thusly determined the individual time evolutions of *Pop1–4* (figure 4) bearing in mind that this time evolution is the convolution of real time evolution of ions flux and the rectangular function corresponding to the acquisition time of 20 µs. Then, we classified the populations in two groups based on their arrival times at the spectrometer; the high-energy populations (*Pop3* and *Pop4*) were detected between 0 and 150 µs after the discharge and the low-energy populations were present during the entire post-discharge period. This distinction implies that *Pop3* and *Pop4* are transported ballistically; they reach the spectrometer aperture just after the end of the discharge, as in the case of energetic neutral (EN) atoms, whereas *Pop1* and *Pop2* exhibit collisional transport more similar to diffusion (see figure 3 in [14]), as in the case of thermalized neutral (TH) atoms. Notably, *Pop3* seems to evolve quite differently from the other three populations. With increasing pressure, the intensities of the other populations grow at the expense of *Pop3*, although we could identify no variation in terms of energy: *Pop3* acts as a reservoir for the other populations via collisions with the background gas.

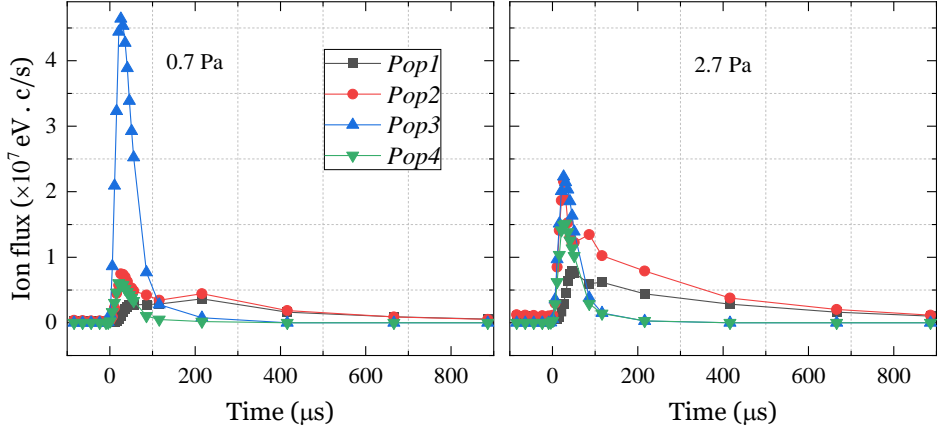

**Figure 4.** Temporal evolution of four Ti$^+$ populations at $P_{HiPIMS}$ = 350 W cm$^{-2}$, $z$ = 5 cm, pure Ar, and $p$ = 0.7 (left) or 2.7 Pa (right).

Avino et al [28] performed such time-resolved measurements at high temporal resolution by selecting only the energies corresponding to the maxima of the distributions of each population. Moreover, Breilmann et al [29] combined such high temporal resolution measurements with high energy resolution measurements. Here, we chose to characterize the IEDFs at high energy resolution and relatively low temporal resolution in order to describe the entire HiPIMS period (1 ms) within a reasonable analysis time. Indeed, our main goal is to better understand the influence of plasma parameters on the evolution of the different ion populations and provide information on their physical origins.

## III.1.2. Reactive sputtering

The introduction of a reactive molecule, here N$_2$, to the argon HiPIMS discharge should generally lead to a variation of the electron density and energy. Due to the vibrational and rotational energies of the N$_2$ molecule, the electrons in the plasma should dissipate their energy by inelastic collisions. Furthermore, the reactive gas contributes to the formation of a compound layer on the target surface, which then affects parameters linked to the surface state, including the secondary electron emission coefficient [30], the sputtering yield, the cathode current, and the cathode voltage at a fixed HiPIMS mean power. In addition, in such a discharge, the sputtered atoms are highly ionized (usually >50%), possibly modifying the electron density [3]. In our previous work, the measurements by means of time average mass spectrometry (for Ti$^+$ ions) and TR-TDLIF (for Ti neutral atoms) revealed a strong decrease of Ti$^+$ and Ti fluxes when 1% N$_2$ is added to gas mixture (see figures 8 and 13 in [14]). For this strong decrease of both species, the poisoning target was identified as the main cause because the surface effect on sputtering dominates that of the plasma in our conditions.

In the present work, we present new TR-IEDFs' measurements to confirm that the fast target poisoning can be observed on each Ti$^+$ population when nitrogen is added in HiPIMS discharge. Figure 5 shows the temporal evolution of the four populations of Ti$^+$ ions at different percentages of nitrogen in the background gas mixture at 2.7 Pa and a distance of 5 cm between the mass spectrometer orifice and the target. As expected, each population's signal decreases strongly when only 1% N$_2$ is added to the gas mixture, although further increasing the fraction of N$_2$ does not significantly affect any population's temporal evolution.

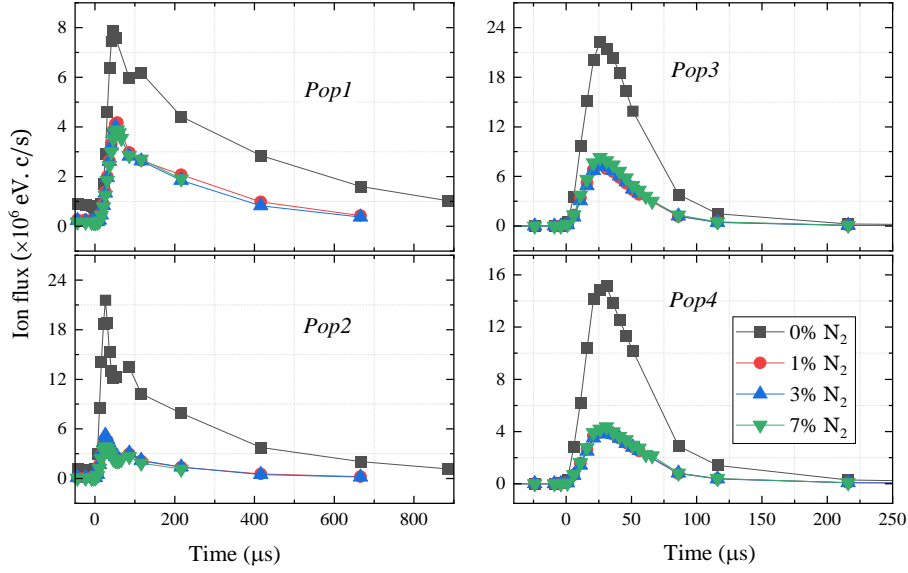

**Figure 5.** Time evolution of four Ti$^+$ populations during the HiPIMS period under different fractions of N$_2$ in the background gas at $P_{\text{HiPIMS}} = 350$ W cm$^{-2}$, $z = 5$ cm, and $p = 2.7$ Pa.

## III.2. Characterization of Ti$^+$, Ti$^{++}$, N$^+$, N$_2^+$, and Ar$^+$

We measured TR-IEDFs of Ti$^{++}$, N$^+$, N$_2^+$, and Ar$^+$ to better characterize the HiPIMS process, and particularly to highlight the types of ions generated in the plasma volume. The major difficulty here is to analyse the distributions of the different types of ions measured by the mass spectrometer because they can be generated at different discharge times and at different locations between the sputtered target and the deposition area (here, the aperture of the mass spectrometer). Therefore, their arrival times do not necessarily correspond to the transport time between the target (or the dense plasma) and the spectrometer. Our main objective in characterizing these ions was therefore to better understand their production; by comparing their TR-IEDFs, we hope to constrain the possible explanations of the four observed Ti$^+$ populations.

Figure 6 shows the TR-IEDFs of Ti$^+$, Ti$^{++}$, and N$^+$ ions with 3% N$_2$ in the gas mixture and at $p = 0.7$ and 2.7 Pa. Four populations coexist for both Ti$^{++}$ and N$^+$, and their energy distributions are similar to those of Ti$^+$, except that the energies of Ti$^{++}$ ions are doubled. The similarity between Ti$^+$ and Ti$^{++}$ is obvious because they are generated by the same initial source: the sputtering of the target. Assuming that titanium atoms are ejected from the target in a neutral state, some are ionized in the high-density plasma to form Ti$^+$, a fraction of which are re-ionized to produce Ti$^{++}$.

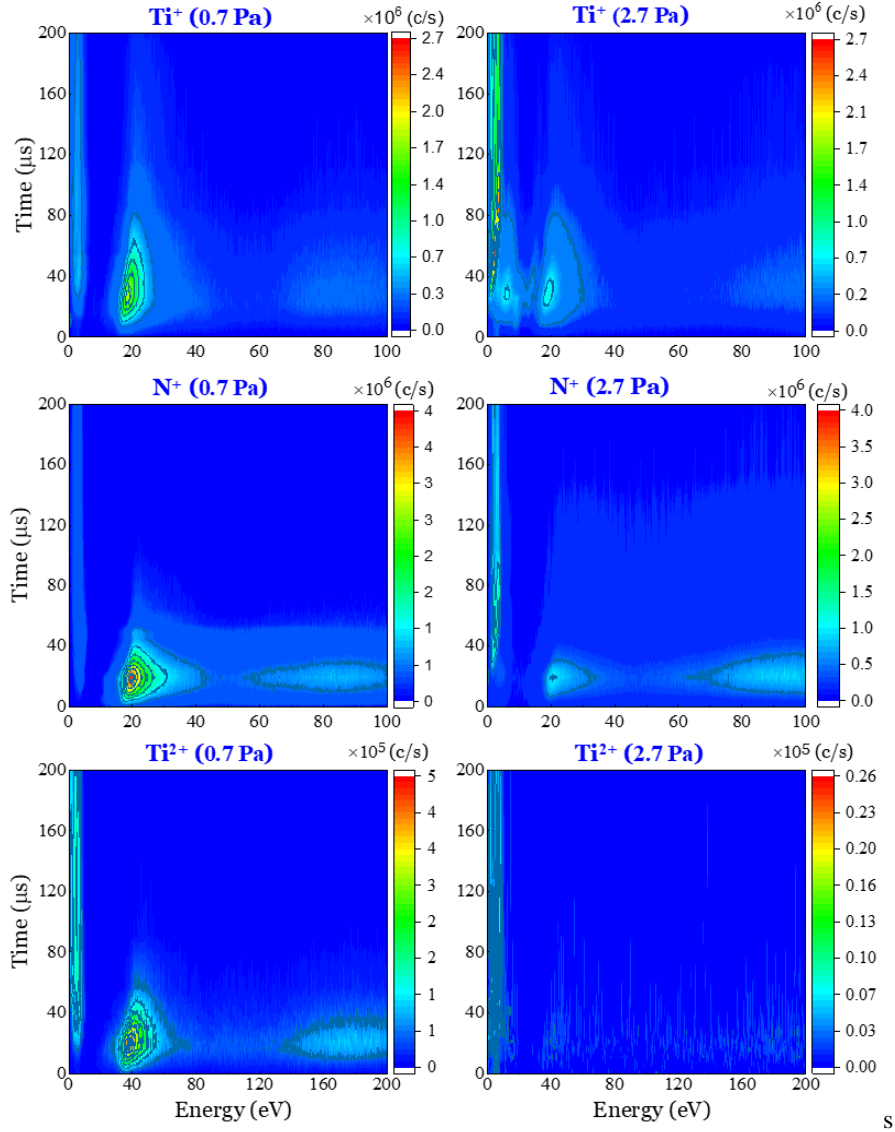

**Figure 6.** TR-IEDFs of (top) Ti$^+$, (middle) N$^+$, and (bottom) Ti$^{++}$ measured at $P_{HiPIMS}$ = 350 W cm$^{-2}$, $p$ = 0.7 (left) or 2.7 Pa (right), 3% N$_2$ in the gas mixture, and $z$ = 5 cm.

The similarity of the N$^+$ and Ti$^+$ distributions suggests that nitrogen ions may be produced by two main processes: dissociation of the N$_2$ molecules injected into the gas mixture and/or sputtering of the nitride target. Indeed, nitrogen atoms present on the surface of the target are sputtered in the same way as titanium atoms and can then be ionized in the plasma during transport. The very strong intensity of *Pop3* in the N$^+$ IEDF, which is predominant at low pressure compared to the other populations, suggests that a significant fraction of the N$^+$ ions come from the sputtering.

Moreover, the number of N$^+$ ions arriving at the aperture (i.e., the ion flux of the time-averaged IEDFs) increases slightly with increasing percentage of N$_2$ in the gas (figure 7). Whereas this increase is expected to be proportional to the increase in the fraction of N$_2$ as seen on the *pop1*, the number of N$^+$ ions only increases by a factor of 1.5 from 1 to 15% N$_2$ in the gas. This seems to confirm that most of the N$^+$ ions are produced by sputtering. The small observed increase must therefore represent the contribution from the plasma due to the dissociation of N$_2$ molecules because the target is quickly saturated with nitrogen at only 1% N$_2$, as was previously reported for R-HiPIMS experiments sputtering a Cr target in an Ar/N$_2$ gas mixture over relatively long discharge times (200 μs) [31,32].

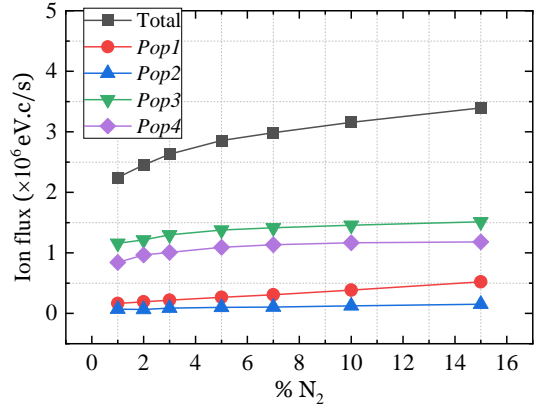

**Figure 7.** N$^+$ population variations as a function of the percentage of N$_2$ in the gas at $z$ = 5 cm, $P_{\text{HiPIMS}}$ = 350 W cm$^{-2}$, and $p$ = 0.7 Pa.

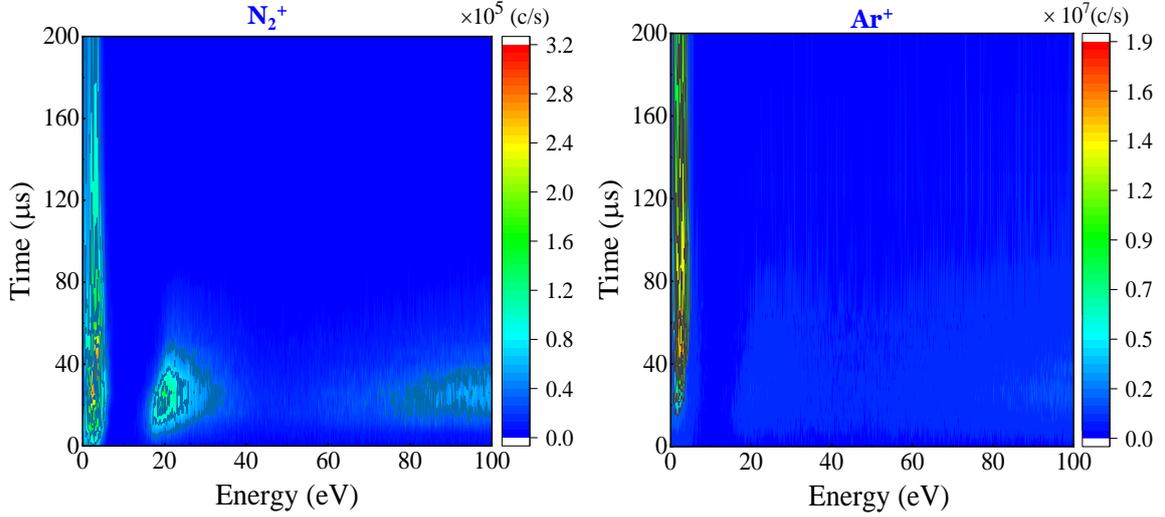

**Figure 8.** TR-IEDFs of Ar$^+$ and N$_2^+$ ions at $z$ = 5 cm, $P_{\text{HiPIMS}}$ = 350 W cm$^{-2}$, $p$ = 0.7 Pa, and 3% N$_2$ in the gas.

For comparison, Ar$^+$ and N$_2^+$ ions issued from the ionization of the neutral species present permanently in the gas mixture have very similar distributions (figure 8). In addition, the TR-IEDFs of these ions are different from those dominantly produced by sputtering. Whereas *Pop3* is dominant at low pressure (0.7 Pa) for Ti$^+$, N$^+$, and Ti$^{2+}$, *Pop1* is the dominant N$_2^+$ and Ar$^+$ population under any discharge conditions. The IEDFs of these gas ions will be used to discuss some processes behind the physical origin of the four populations.

## III.3. The physical origin of four IEDF populations

We here discuss the physical origins of the populations highlighted in section III.1 by relying on our mass spectrometry, voltage probe, and TR-TDLIF measurements. To better understand the origin of each population, we discuss them in the following order: *Pop3, Pop2, Pop1,* and finally *Pop4*.

### III.3.1. Population 3 (15–50 eV)

The physical origin of *Pop3* has been the subject of several previous works [25,26,33,34]; this population is the most complicated population to explain, especially in terms of energy, and several explanations have been proposed. One possible explanation is that *Pop3* corresponds to the ionization of high-energy neutral atoms (the tail of the Thompson distribution [27]). Fast camera footage revealed the formation of structures called 'spokes' that turn above the target in the azimuthal direction ($\vec{E} \times \vec{B}$) [21,22], although the mechanisms generating these structures and their influence on ionic transport properties are not yet well understood. Anders et al [35] measured the IEDF of $Nb^+$ ions in a plane close to the target by mass spectrometry in two different geometries. By orienting the axis of the mass spectrometer parallel to the target surface such that $\vec{E} \times \vec{B}$ pointed towards or away from the detector (i.e., in the $r$ or $-r$ direction, respectively), they showed that higher energy ions are preferentially transported in the azimuthal direction ($r$). They proposed that a 'double layer' configuration, i.e., an azimuthally asymmetric shape of the potential in the spokes along the racetrack, could explain the acceleration of the charged particles in all directions (including the axial direction) and the asymmetry of the IEDFs, especially for high-energy ions (here, *Pop3*, 15–50 eV). This third population has also been attributed to fast ions reflecting off of the target [17].

Figure 9 shows the IEDFs of $Ti^+$, $N^+$, $Ti^{++}$, $Ar^+$, and $N_2^+$ at $t = 30$ µs for two pressures ($p = 0.4$ and 2.7 Pa), and two distances from the target ($z = 5$ and 12 cm). Except at $z = 12$ cm and $p = 2.7$ Pa, *Pop3* is detected for all types of ions, especially at low pressure (0.4 Pa). However, Maszl et al [26] did not detect *Pop3* in gas species in a HiPIMS discharge with a relatively long discharge duration of 100 µs. If this population were related to the sputtering (Ti and N), it should not be detected for $Ar^+$ and $N_2^+$. Moreover, the average energy of ballistic neutral sputtered titanium atoms decreases during the post-discharge period; previous works using the same plasma conditions found that the average energy of the energetic Ti population decreases from 8 to about 1 eV [14]. This decrease of the average energy of ballistic atoms is not due to collisions. By adopting a short discharge time to clearly separate the sputtering and transport phases, our experiments show that the high-energy atoms (8 eV) crossed the probed volume by $t = 5$ µs, corresponding to the TOF of the first high-energy sputtered atoms between the target surface and the probed volume. Similarly, the low-energy atoms (1 eV) crossed the probed volume by $t = 25$ µs, corresponding to the TOF of the last low-energy sputtered atoms. Therefore, the decrease in the average energy of ballistic atoms is mainly due to sputtering. Because this behaviour is not observed for the average energy of *Pop3* of $Ti^+$, which remains almost constant during the post-discharge period at ~19 eV for singly charged ions and ~38 eV for doubly charged ions, and the peaks of *Pop3* of $Ti^+$, $N^+$, $Ti^{++}$, $N_2^+$, and $Ar^+$ can only be explained by the acceleration of these ions in the main plasma that produces them.

Mishra et al [36] used an electron-emitting probe to measure the temporal evolution of the plasma potential along the axial direction ($z$) above the racetrack. They reported a strongly negative plasma potential gradient in the ionization region (the region where the electrons are magnetically trapped) at the beginning of the discharge (in the first 6–8 µs). After 55–60 µs, the plasma potential gradient becomes weak and the bulk electric field collapses. Notably, the potential decreases towards the target. The same behaviour was reported by Revel et al [37] in 2D particle-in-cell Monte Carlo collision simulations of HiPIMS plasma with short discharge duration. Thus, if the *Pop3* ions were produced in the ionization region, they should be attracted by the voltage decrease towards the cathode instead of accelerated towards the grounded orifice of the mass spectrometer. We note that these studies were not adapted to predict the effects of spokes on the plasma potential.

As mentioned above, the dense HiPIMS plasma near the target is not homogeneous, and can exhibit one or more spokes with a rotational velocity of about 10 km s$^{-1}$ in the $\vec{E} \times \vec{B}$ azimuthal direction. Recently, Held et al [38] characterized the plasma potential inside those structures using time-shift averaging Langmuir probe measurements. They reported that the plasma potential becomes higher inside a spoke than in the surrounding region, with the measured potential in those structures being positive and having a weak gradient. In standard measurements, i.e., without time-shift averaging, they found that the plasma potential in the ionisation region becomes more negative when the measurements are performed close to the target, as also reported elsewhere [36,37]. The plasma potential fluctuations induced by the presence of a spoke, from a minimum in the surrounding region to a maximum inside the spoke, are of about 7 V [38]. Anders et al [35], who proposed the double-layer configuration, predicted potential fluctuations of ~35 V for HiPIMS spokes and suggested that the electric field induced by a spoke in the magnetic trap region could increase the ion flux towards the substrate.

Our measurements show that *Pop3* of singly charged ions has an energy centred at 19 eV and doubly charged ions has an energy centred at 38 eV, regardless of the plasma conditions (post-discharge time, pressure, gas mixture). These above findings therefore suggest that only spokes and their potential structures can generate such ion energies towards the substrate. The energy of $Ti^{2+}$ strongly supports the argument that pop3 ions are accelerated by plasma potential inside the spokes. Based on these considerations, the average energy of those ions suggests that the potential fluctuations inside the spokes are of about 19 V under our plasma conditions. The flux of *Pop3* (the mass spectrometer signal) thus depends on the gas pressure and the distance from the target.

Three effects can modify the *Pop3* signal: the acceptance angle of the mass spectrometer, a geometric effect, and thermalization with the background gas (a pressure effect).

- Acceptance angle. As described in the experimental setup, the mass spectrometer orifice is in front of the centre of the target. The sputtering occurs mainly in the racetrack, and the ionization of sputtered particles occurs in the ionization region. An ion produced in this region arrives at the mass spectrometer orifice at an angle $\theta$, where $\tan(\theta) \approx \frac{R_0}{z-z_0}$, $R_0 = 1.25$ cm, and $z_0 \approx 1$ cm (figure 10). For $z = 5$ and 12 cm, $\theta$ is 16° and 6.1°, respectively. Thus, with increasing distance $z$, the likelihood that these ions are measured by the mass spectrometer increases [39].
- Geometrical effect. The *Pop3* ions are produced in the ionization region above the racetrack and, ideally, transported uniformly in all directions. Therefore, *Pop3* ions spread in a sphere and their flux should decrease as the area of the sphere ($S = 4\pi \times [R_0^2 + (z - z_0)^2)]$ increases (figure 10).
- Pressure effect. When the product of pressure × distance is increased, ions undergo more collisions with the buffer gas, inducing a decrease of the *Pop3* signal.

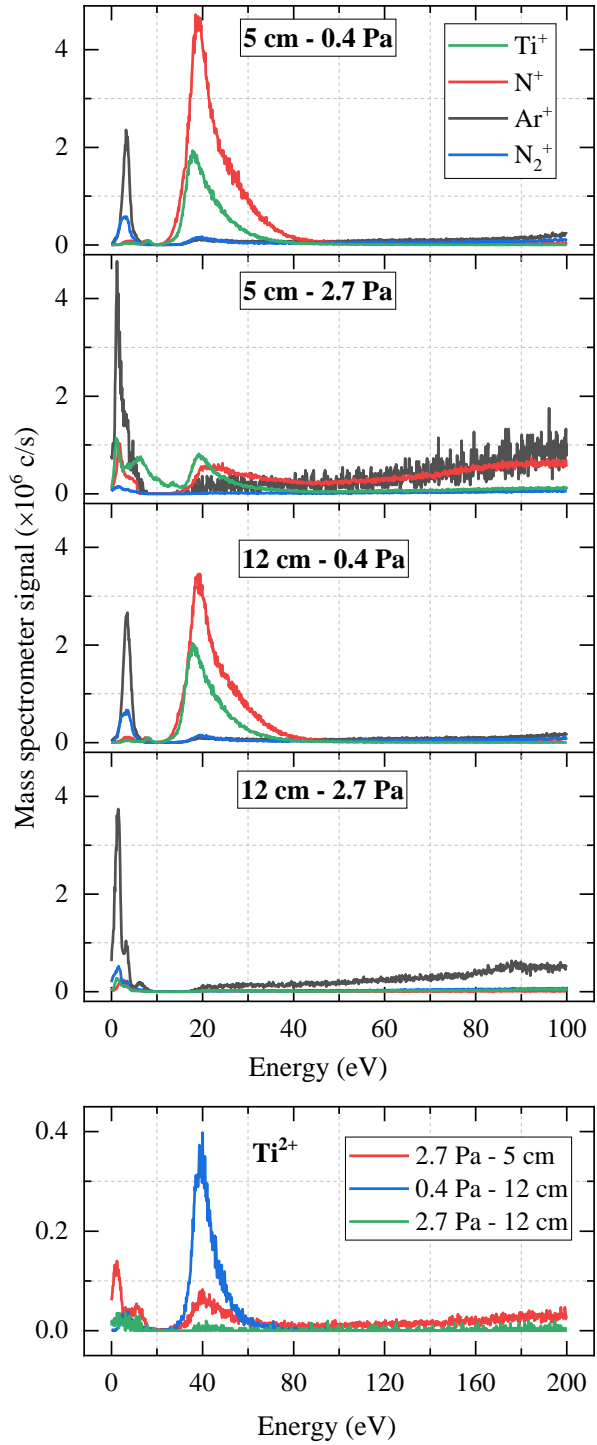

**Figure 9.** TR-IEDFs of $Ar^+$, $N_2^+$, $Ti^+$, and $N^+$ (top panels) and $Ti^{++}$ (bottom panel) at $t = 30$ μs, $P_{HiPIMS} = 350$ W cm$^{-2}$, 3% $N_2$ in the Ar gas, $p = 0.4$ and 2.7 Pa, and $z = 5$ and 12 cm from the target.

At low pressure (0.4 Pa), i.e., when collisions between particles are limited, our time-resolved measurements of $Ti^+$, $N^+$, and $Ti^{++}$ show that *Pop3* is the dominant population, whatever the distance from the target ($z = 5$ and 12 cm) (figure 10). At higher pressure (2.7 Pa), the signals of *Pop1*, *Pop2*, and *Pop3* are comparable at $z = 5$ cm, whereas the signal of *Pop3* decreases drastically at $z = 12$ cm. Regarding $Ar^+$ and $N_2^+$, the contribution of *Pop1* clearly dominates, whatever the measurement conditions. These results indicate that *Pop3* ions are highly affected by thermalization during collisions with the gas particles and by the spherical area over which they spread. Moreover, the acceptance angle of the mass spectrometer (figure 10) does not significantly alter the *Pop3* signal intensity in our experiments, although this effect is well observed closer to the target ($z = 2$–$4$ cm).

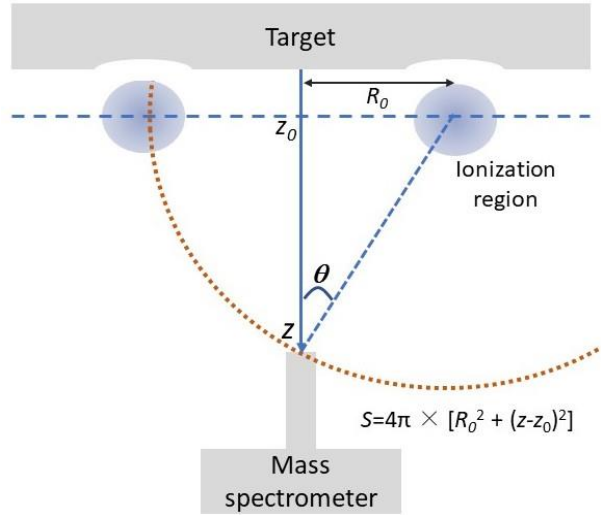

**Figure 10.** A schematic cross-section of the ionization region, target, and mass spectrometer. $\theta$ is the angle at which an ion produced in the ionization region is measured by the mass spectrometer. *S* is the spherical area over which the *Pop3* ions spread.

## III.3.2. Population 2 (4–15 eV)

Although we detected ions of *Pop2* at 0–15 eV, it is difficult to distinguish them from *Pop1* due to the overlap of these two populations between 0 and 4 eV. Therefore, we restricted our analysis of *Pop2* to 4–15 eV, as shown in figure 3. The production of *Pop2* ions is greater under higher discharge pressure (2.7 Pa; top panels of figure 11), and the average energy of *Pop2* decreases with increasing pressure and distance from the target (bottom panels of figure 11). Because the intensity of *Pop2* increases with discharge pressure whereas that of *Pop3* decreases (figure 4), and because it is well known that the collisional process depends on pressure and the distance from the target, a part of *Pop3* and *Pop4* must feed some portion of *Pop2* through collisions with the gas.

*Pop2* is intermediate between *Pop3* and *Pop1*, which is equivalent to the quasi-thermalized atoms (QTH) detected in TR-TDLIF measurements of neutral Ti atoms [14]. A second process could therefore also contribute to the generation of *Pop2*. This process concerns the recombination of $Ti^{++}$ ions with electrons to become $Ti^+$ or the charge exchange between $Ti^{++}$ and argon atoms, as discussed by Hecimovic and Ehiasarian [27] for $Cr^{++}$ and $Cr^+$ ions. Nonetheless, the ionization potential of $Ti^+$ is 13.57 eV and that of Ar atoms is 15.76 eV. Therefore, the energy defect for the charge exchange between these species is $-2.2$ eV, and this process can be ruled out. Comparing the IEDFs of titanium in figure 9, it is apparent that *Pop2* of $Ti^+$ and $Ti^{++}$ have the same energy (between 4 and 15 eV), whereas *Pop3* and *Pop4* of $Ti^{++}$ are twice as energetic those of $Ti^+$. This indicates that the recombination $Ti^+$ with electrons directly feeds *Pop2* of $Ti^+$ in the 0–15 eV energy range, even if no $Ti^{++}$ ions have energies of 15–30 eV. Nevertheless, this process seems minor under our conditions, and the signal of *Pop2* of $Ti^{++}$ ions is a decade smaller than that of $Ti^+$ ions.

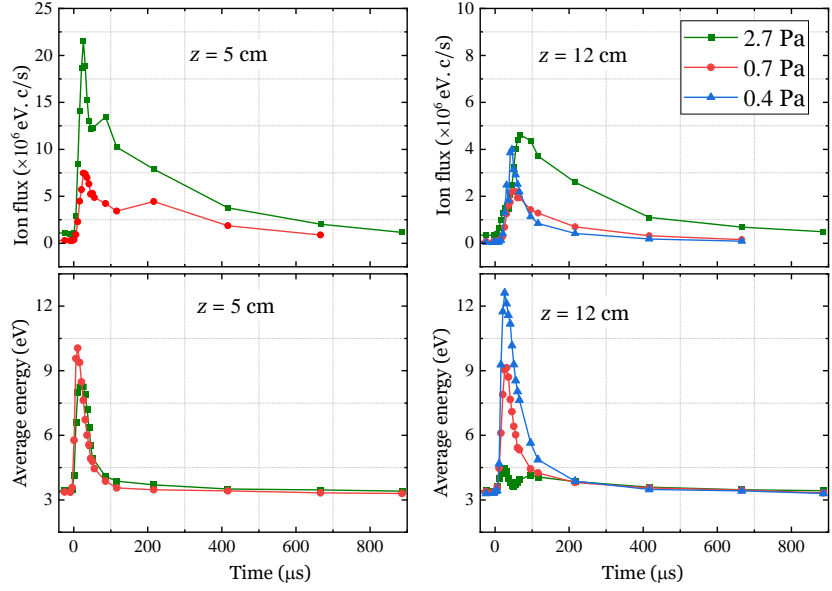

**Figure 11.** The temporal evolution of the ion flux (equation 1, upper panels) and average energy (bottom panels) of *Pop2* of Ti$^+$ at different pressures and $z = 5$ cm (left) and 12 cm (right), $P_{\text{HiPIMS}} = 350$ W cm$^{-2}$, and in a pure Ar gas.

### III.3.3. Population 1, thermalized ions (0–4 eV)

Because *Pop1* is detectable during the entire post-discharge period (1 ms), there are two possible processes that could lead to its generation: the ionization of thermalized neutral (TH) atoms or the thermalization of fast ions from other, more energetic populations (*Pop2–4*) during transport.

At 2.7 Pa, two peaks are observed in the signal of *Pop1* of Ti$^+$ (figure 12): the first at $t \sim 40$ µs, just after the discharge, and the second shortly after, at $t \sim 116$ µs. At lower pressures (0.4 and 0.7 Pa), the first peak does not appear. To better understand the process producing ions of *Pop1*, figure 13a compares the Ti$^+$ IEDF at $t = 36$ µs with its corresponding fit obtained using a shifted Maxwellian distribution function [29]

$$f(E) \propto g(E) \exp\left(-\frac{(\sqrt{E}-\sqrt{E_{\text{shift}}})^2}{T_{\text{ions}}}\right) \quad (2)$$

where $T_{\text{ions}}$ is the ion temperature and $E_{\text{shift}}$ mainly represents the energy gained by ions in the sheath in front of the mass spectrometer. Because the orifice of the mass spectrometer is grounded in our experiments, $E_{\text{shift}}$ represents the plasma potential, assuming a collisionless sheath. The prefactor $g(E)$ depends on the acceptance angle of the mass spectrometer, which should be constant in our experiments [29]. At $t = 36$ µs, the fit to our data indicates that the ion temperature of *Pop1* is around 1100 K and that *Pop1* is shifted ($E_{\text{Shift}}$) by about 2.5 eV. Given that the mass spectrometer orifice is grounded, this energy shift indicates that the plasma potential at $z = 5$ cm is about +2.5 V, consistent with previous results [36,37]. In the velocity distribution function of neutral Ti atoms measured previously by TR-TDLIF (figure 13b), three groups of atoms were identified: TH, QTH, and EN atoms [14]. TH atoms have a Maxwellian velocity distribution centred at an axial velocity of 0 m s$^{-1}$ and a temperature around 800 K. At $t = 36$ µs, $T_{\text{ions}}$ of *Pop1* and the temperature of TH atoms are similar. One could estimate that the 300-K discrepancy between the temperatures represents the errors on the measurement and the fit. The error seems generated mainly from the variation of plasma potential during the acquisition time of TR-IEDFs in which was a gate of 20 µs. The *Pop1* peak of Ar$^+$ at $t = t = 36$ µs was also fitted with the shifted

Maxwellian and the obtained the temperature was 1000 K. This suggests that *Pop1* ions and TH atoms are in local thermal equilibrium at 2.7 Pa, 5 cm from the target, and $t \geq 36$ μs.

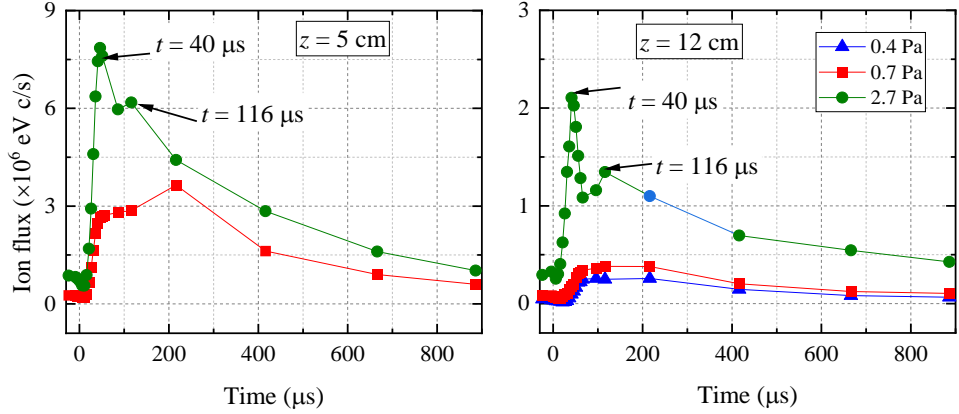

**Figure 12.** The temporal evolution of the ion flux (equation 1) of *Pop1* of Ti$^+$ (0–4 eV) at different pressures and $z = 5$ cm (left) and 12 cm (right), $P_{\text{HiPIMS}} = 350$ W cm$^{-2}$, and in pure Ar gas.

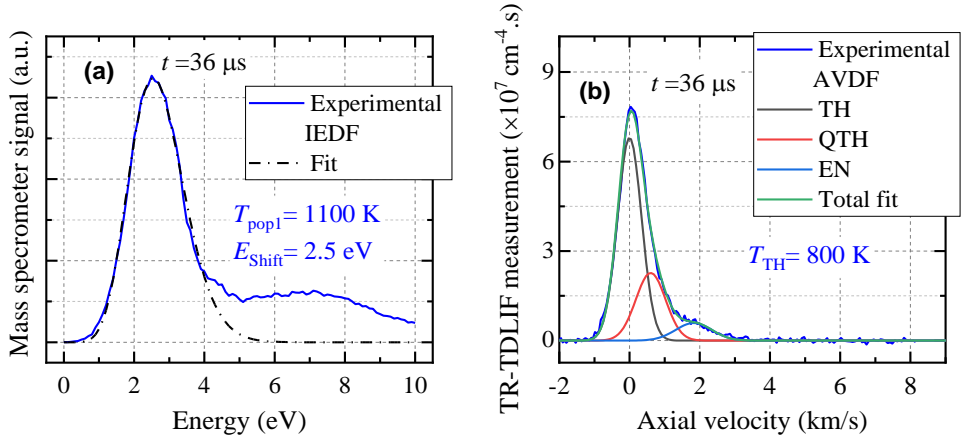

**Figure 13.** (a) IEDF of Ti$^+$ ions at 36 μs, and the corresponding fit of *Pop1* by a shifted Maxwellian distribution (equation 2). (b) Neutral Ti velocity distribution function measured by TR-TDLIF at $t = 36$ μs. Discharge conditions: $p = 2.7$ Pa, $z = 5$ cm, $P_{\text{HiPIMS}} = 350$ W cm$^{-2}$, and pure Ar gas.

In addition, the temporal evolution of the flux of thermalized neutral Ti atoms ($\Phi_{\text{TH}}$) toward the substrate (i.e. we took only atoms with positive velocity in z-direction) at $p = 2.7$ Pa is plotted in figure 14. At the end of the post-discharge period ($t = 1000$ μs), residual thermalized neutral atoms remain in the measurement volume and are detected again at the beginning of the next discharge ($\Phi_{\text{TH}} (t = 0$ μs$) = \Phi_{\text{TH}} (t = 1000$ μs$)$; black dashed line for $z = 5$ cm in figure 14). The transport of TH atoms is diffusive, and any losses caused by this process are very weak at the relatively high magnetron pressure of 2.7 Pa. At the beginning of the pulse, $\Phi_{\text{TH}}$ decreases faster than at the end of the post-discharge period, reaching a minimum value at $t = 30$ μs, when the newly produced TH atoms are observed in the probed volume. We note that this time is close to that corresponding to the maximum of the first peak of *Pop1* of Ti$^+$.

These results seem to show that *Pop1* ions are produced by the very first Ti$^+$ ion generation process: the ionization of residual thermal neutral atoms remaining from the previous post-discharge period. Indeed, at lower pressure, much fewer

TH atoms are generated by collisions with the buffer gas, and their diffusive transport (loss) is faster. Therefore, at 0.7 and 0.4 Pa, fewer residual TH atoms remain at the end of the post-discharge period, and their ionization by the subsequent discharge (the first peak of *Pop1*) is less obvious.

The second peak in the temporal evolution of *Pop1* (figure 12) may simply be a low-energy portion of *Pop2*; it appears ~100 μs after the end of the discharge, when the electron energy is not enough to efficiently ionize the thermalized vapor. Therefore, these ions could correspond to the thermalization of energetic ions, as discussed in section III.3.2.

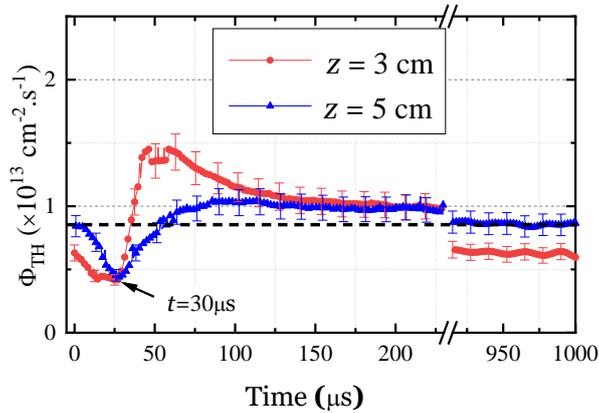

**Figure 14.** Temporal evolution of the flux of thermalized neutral titanium atoms in the *z*-direction toward the substrate (see figure 1) measured by TR-TDLIF [14] at different distances from the target, $p$ = 2.7 Pa, $R_0$ = 1.3 cm, $P_{HiPIMS}$ = 350 W cm$^{-2}$, and in pure Ar gas. The corresponding density of thermalized atoms at these plasma conditions is in the order of $10^9$ cm$^{-3}$.

## III.3.4. Population 4 (50–100 eV)

To better understand the physical origin of the population of very highly energetic ions, we measured the temporal evolution of the cathode voltage using a high-voltage probe with a large bandwidth. A measurement at 2.7 Pa and in pure Ar gas is shown in figure 15. At the end of the discharge, four peaks of positive voltage appear between 12 and 14 μs. These peaks are real, and are not due to the electrical noise generated by the high power (as at the beginning of the discharge). Indeed, if they were simply electrical noise, they would appear as soon as the discharge was turned off (i.e., at 10 μs) and oscillate around the grounded voltage reference, alternating between positive and negative values. These positive peaks in the cathode voltage are thus capable of repelling positive ions and attracting electrons. This characteristic is certainly specific to our generator and is probably how the manufacturer allowed working with short pulses, i.e., by pushing positive charges near the cathode. We note that these peaks disappear when the discharge duration is increased.

Keraudy et al. [40] studied the ion energy distribution in bipolar HiPIMS discharges, in which a positive voltage pulse follows the high-power impulse to explore the influence of a long positive-voltage pulse (200 μs) in the range 10–150 V. Their mass spectrometry results clearly showed that a large proportion of metal ions is accelerated with an energy that corresponds to the positive voltage. Thus, they interpreted that those ions came from the region very close to the target, where the ions were repelled by the positive cathode voltage.

*Pop4* is well observed at ~2.7 Pa, but is attenuated at lower pressure (~0.4 Pa; figure 16). In addition, the second and third positive-voltage peaks (~100 V in the inset of figure 15) are very close to the energy of *Pop4*, about 90 eV (top panel, figure 16). This small 10-eV difference could be explained by thermalization during ion transport into the mass spectrometer. Furthermore, this population is not observed in DC discharges. These findings seem to confirm that the ions of *Pop4* are accelerated by the positive cathode voltage.

Finally, the variations of *Pop4* of Ti$^+$ ions as functions of pressure and distance (figure 17) are similar to those observed for *Pop1* and *Pop2*. In fact, the generation of *Pop4*, especially metallic ions, is favoured when the operating conditions allow the production of enough thermalized ions in the vicinity of the target; if the operating conditions can only produce *Pop3* ions, they cannot be accelerated by the positive cathode voltage because they are already moving away very quickly at the beginning of the post-discharge period. It thus seems that *Pop4* is mainly due to the acceleration of thermalized ions, namely those of *Pop1*. This may explain why *Pop4* ions are observed in the TR-IEDFs of N$_2^+$ and Ar$^+$ even at 0.4 Pa (figure 8).

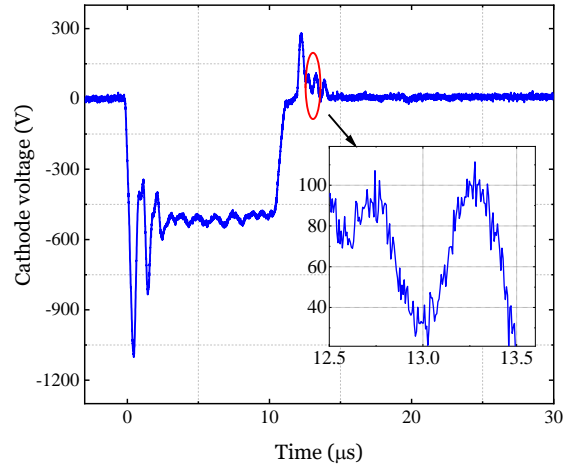

**Figure 15.** Evolution of the cathode voltage at $P_{\text{HiPIMS}} = 350$ W cm$^{-2}$, $p = 2.7$ Pa, and in pure Ar gas.

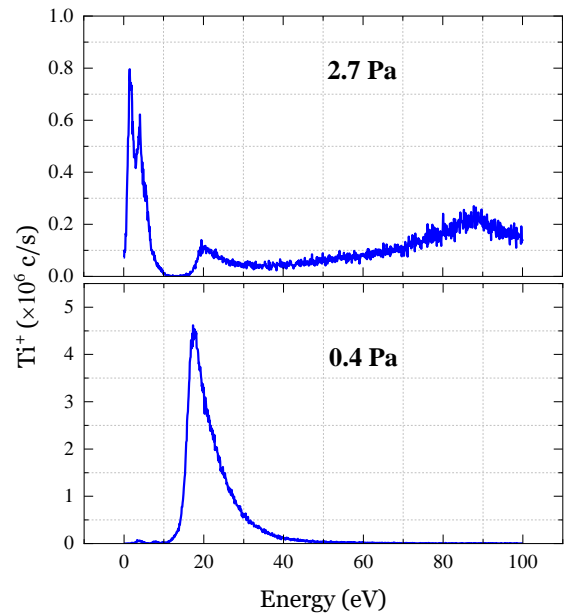

**Figure 16.** TR-IEDFs of Ti$^+$ ions at 2.7 and 0.4 Pa, $z = 12$ cm from the target, $t = 30$ μs, $P_{\text{HiPIMS}} = 350$ W cm$^{-2}$, and in pure Ar gas.

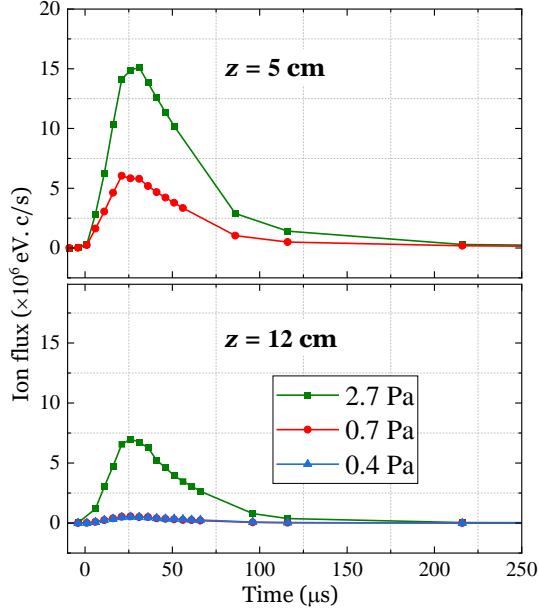

**Figure 17.** Temporal evolution of the ion flux of *Pop4* of Ti$^+$ ions between 50 and 100 eV (see equation 1) at different pressures, $z = 5$ and 12 cm, $P_{HiPIMS} = 350$ W cm$^{-2}$, and in pure Ar gas.

## IV. Conclusions

This work is mainly devoted to characterizing the transport of metal ions (Ti$^+$ and Ti$^{++}$) and ions produced from the buffer gas (N$^+$, Ar$^+$, and N$_2^+$) during the HiPIMS process. Time-resolved mass spectrometry measurements of ion energy distribution functions (TR-IEDFs) revealed four populations (*Pop1, Pop2, Pop3,* and *Pop4*) that vary depending on the working conditions (pressure, percentage of nitrogen in the Ar/N$_2$ gas mixture, and distance from the target). The variations of Ti$^+$ populations as a function of the percentage of N$_2$ reveal the same behaviour observed for neutral Ti atoms, i.e., a strong decrease in the flux when only 1% N$_2$ is added to the argon gas, but no variation at greater N$_2$ percentages. Thus, at >1% N$_2$, the percentage of N$_2$ in the reactive gas does not modify the production of metallic ions. This behaviour can therefore be explained by our working conditions, particularly the relatively low HiPIMS power (350 W cm$^{-2}$) chosen for the TR-TDLIF measurements and the short discharge time (10 μs), which does not allow for the sufficient removal of the nitride layer formed on the target surface.

Our comparison and analysis of the IEDFs of Ti$^+$, Ti$^{++}$, and N$^+$ as a function of the fraction of N$_2$ in the gas indicate that most of the neutral N atoms are produced by the sputtering of the nitride layer on the target, with just a small portion being produced by the dissociation of nitrogen molecules in the plasma.

We explored the physical origins of the four observed ion populations. *Pop1* is produced by the ionization of thermalized neutral atoms. *Pop3* is the original ion population; at low pressures where collisions with the gas are limited, only *Pop3* can be detected by mass spectrometry. This population thus seems to be generated by spokes. *Pop2* is mainly generated by collisions between the ions of *Pop3* (and probably *Pop4*) and the buffer gas during transport, but may also result from the recombination of electrons and doubly ionized ions. *Pop4* seems to be produced by the specific features of our HiPIMS generator; positive peaks in the cathode voltage that appear when short-duration HiPIMS discharges are turned off accelerate ions of *Pop1* (and probably *Pop2*) present in the vicinity of the cathode, similar to a bipolar HiPIMS process.

# References

[1]   Kouznetsov V, Macák K, Schneider J M, Helmersson U and Petrov I 1999 A novel pulsed magnetron sputter technique utilizing very high target power densities *Surface and Coatings Technology* **122** 290–3

[2]   Anders A 2017 Tutorial: Reactive high power impulse magnetron sputtering (R-HiPIMS) *Journal of Applied Physics* **121** 171101

[3]   Sarakinos K, Alami J and Konstantinidis S 2010 High power pulsed magnetron sputtering: A review on scientific and engineering state of the art *Surface and Coatings Technology* **204** 1661–84

[4]   Ehiasarian A P, Vetushka A, Gonzalvo Y A, Sáfrán G, Székely L and Barna P B 2011 Influence of high power impulse magnetron sputtering plasma ionization on the microstructure of TiN thin films *Journal of Applied Physics* **109** 104314

[5]   Somekh R E 1984 The thermalization of energetic atoms during the sputtering process *Journal of Vacuum Science & Technology A* **2** 1285–91

[6]   Desecures M, Poucques L de and Bougdira J 2014 Characterization of energetic and thermalized sputtered tungsten atoms using tuneable diode-laser induced fluorescence in direct current magnetron discharge *Plasma Sources Science and Technology* **24** 015012

[7]   Held J, Hecimovic A, von Keudell A and Schulz-von der Gathen V 2018 Velocity distribution of titanium neutrals in the target region of high power impulse magnetron sputtering discharges *Plasma Sources Science and Technology* **27** 105012

[8]   Britun N, Palmucci M and Snyders R 2011 Fast relaxation of the velocity distribution function of neutral and ionized species in high-power impulse magnetron sputtering *Applied Physics Letters* **99** 131504

[9]   Shibagaki K, Nafarizal N and Sasaki K 2005 Spatial distribution of the velocity distribution function of Fe atoms in a magnetron sputtering plasma source *Journal of Applied Physics* **98** 043310

[10]  Sushkov V, Do H T, Cada M, Hubicka Z and Hippler R 2012 Time-resolved tunable diode laser absorption spectroscopy of excited argon and ground-state titanium atoms in pulsed magnetron discharges *Plasma Sources Science and Technology* **22** 015002

[11]  Poucques L de, Vitelaru C, Minea T M, Bretagne J and Popa G 2008 On the anisotropy and thermalization of the metal sputtered atoms in a low-pressure magnetron discharge *EPL (Europhysics Letters)* **82** 15002

[12]  Desecures M, de Poucques L, Easwarakhanthan T and Bougdira J 2014 Characterization of energetic and thermalized sputtered atoms in pulsed plasma using time-resolved tunable diode-laser induced fluorescence *Applied Physics Letters* **105** 181120

[13]  Vitelaru C, de Poucques L, Minea T M and Popa G 2011 Space-resolved velocity and flux distributions of sputtered Ti atoms in a planar circular magnetron discharge *Plasma Sources Science and Technology* **20** 045020

[14]  Farsy A E, Ledig J, Desecures M, Bougdira J and Poucques L de 2019 Characterization of transport of titanium neutral atoms sputtered in Ar and Ar/N2 HIPIMS discharges *Plasma Sources Sci. Technol.* **28** 035005

[15]     Desecures M, de Poucques L and Bougdira J 2016 Determination of deposited flux and energy of sputtered tungsten atoms on every stages of transport in HiPIMS discharge *Plasma Sources Science and Technology* **26** 025003

[16]     Ferrec A, Kéraudy J and Jouan P-Y 2016 Mass spectrometry analyzes to highlight differences between short and long HiPIMS discharges *Applied Surface Science* **390** 497–505

[17]     Palmucci M, Britun N, Silva T, Snyders R and Konstantinidis S 2013 Mass spectrometry diagnostics of short-pulsed HiPIMS discharges *Journal of Physics D: Applied Physics* **46** 215201

[18]     Thompson M W 1968 II. The energy spectrum of ejected atoms during the high energy sputtering of gold *Philosophical Magazine* **18** 377–414

[19]     Stepanova M and Dew S 2004 Anisotropic energies of sputtered atoms under oblique ion incidence *Nuclear Instruments and Methods in Physics Research Section B: Beam Interactions with Materials and Atoms* **215** 357–65

[20]     Anders A 2012 Self-organization and self-limitation in high power impulse magnetron sputtering *Applied Physics Letters* **100** 224104

[21]     Anders A, Ni P and Rauch A 2012 Drifting localization of ionization runaway: Unraveling the nature of anomalous transport in high power impulse magnetron sputtering *Journal of Applied Physics* **111** 053304

[22]     Ehiasarian A P, Hecimovic A, de los Arcos T, New R, Schulz-von der Gathen V, Böke M and Winter J 2012 High power impulse magnetron sputtering discharges: Instabilities and plasma self-organization *Applied Physics Letters* **100** 114101

[23]     Hecimovic A and Keudell A von 2018 Spokes in high power impulse magnetron sputtering plasmas *J. Phys. D: Appl. Phys.* **51** 453001

[24]     Bohlmark J, Lattemann M, Gudmundsson J T, Ehiasarian A P, Aranda Gonzalvo Y, Brenning N and Helmersson U 2006 The ion energy distributions and ion flux composition from a high power impulse magnetron sputtering discharge *Thin Solid Films* **515** 1522–6

[25]     Franz R, Clavero C, Kolbeck J and Anders A 2016 Influence of ionisation zone motion in high power impulse magnetron sputtering on angular ion flux and NbO$_x$ film growth *Plasma Sources Science and Technology* **25** 015022

[26]     Maszl C, Breilmann W, Benedikt J and von Keudell A 2014 Origin of the energetic ions at the substrate generated during high power pulsed magnetron sputtering of titanium *Journal of Physics D: Applied Physics* **47** 224002

[27]     Hecimovic A and Ehiasarian A P 2009 Time evolution of ion energies in HIPIMS of chromium plasma discharge *J. Phys. D: Appl. Phys.* **42** 135209

[28]     Avino F, Sublet A and Taborelli M 2019 Evidence of ion energy distribution shift in HiPIMS plasmas with positive pulse *Plasma Sources Sci. Technol.* **28** 01LT03

[29]     Breilmann W, Maszl C, Benedikt J and Keudell A von 2013 Dynamic of the growth flux at the substrate during high-power pulsed magnetron sputtering (HiPIMS) of titanium *J. Phys. D: Appl. Phys.* **46** 485204

[30] Depla D, Mahieu S and De Gryse R 2009 Magnetron sputter deposition: Linking discharge voltage with target properties *Thin Solid Films* **517** 2825–39

[31] Greczynski G and Hultman L 2010 Time and energy resolved ion mass spectroscopy studies of the ion flux during high power pulsed magnetron sputtering of Cr in Ar and Ar/N2 atmospheres *Vacuum* **84** 1159–70

[32] Greczynski G, Lu J, Johansson M P, Jensen J, Petrov I, Greene J E and Hultman L 2012 Role of Tin+ and Aln+ ion irradiation (n=1, 2) during Ti1-xAlxN alloy film growth in a hybrid HIPIMS/magnetron mode *Surface and Coatings Technology* **206** 4202–11

[33] Brenning N, Lundin D, Minea T, Costin C and Vitelaru C 2013 Spokes and charged particle transport in HiPIMS magnetrons *Journal of Physics D: Applied Physics* **46** 084005

[34] Yang Y, Tanaka K, Liu J and Anders A 2015 Ion energies in high power impulse magnetron sputtering with and without localized ionization zones *Applied Physics Letters* **106** 124102

[35] Anders A, Panjan M, Franz R, Andersson J and Ni P 2013 Drifting potential humps in ionization zones: The "propeller blades" of high power impulse magnetron sputtering *Applied Physics Letters* **103** 144103

[36] Mishra A, Kelly P J and Bradley J W 2010 The evolution of the plasma potential in a HiPIMS discharge and its relationship to deposition rate *Plasma Sources Sci. Technol.* **19** 045014

[37] Revel A, Minea T and Costin C 2018 2D PIC-MCC simulations of magnetron plasma in HiPIMS regime with external circuit *Plasma Sources Sci. Technol.* **27** 105009

[38] Held J, Maaß P A, Gathen V S der and Keudell A von 2020 Electron density, temperature and the potential structure of spokes in HiPIMS *Plasma Sources Sci. Technol.* **29** 025006

[39] Lundin D, Larsson P, Wallin E, Lattemann M, Brenning N and Helmersson U 2008 Cross-field ion transport during high power impulse magnetron sputtering *Plasma Sources Sci. Technol.* **17** 035021

[40] Keraudy J, Viloan R P B, Raadu M A, Brenning N, Lundin D and Helmersson U 2019 Bipolar HiPIMS for tailoring ion energies in thin film deposition *Surface and Coatings Technology* **359** 433–7